\documentclass[twocolumn,secnumarabic,amssymb, nobibnotes, aps, pre, floatfix,nofootinbib]{revtex4-2}

\usepackage[english]{babel}
\usepackage[utf8]{inputenc}
\usepackage{amsmath, amssymb, amsthm}
\usepackage{graphics}
\usepackage{physics}
\usepackage{graphicx}
\usepackage{mathrsfs}
\usepackage{blindtext}
\usepackage{verbatim}
\usepackage{lipsum}
\usepackage[hidelinks]{hyperref}
\usepackage{cleveref}
\usepackage{mathdots}

\usepackage{bookmark}

\usepackage{tabularx}
\usepackage{float}

\usepackage[caption=false]{subfig}
\usepackage{tikz}
\usetikzlibrary{positioning,decorations.pathreplacing}

{%TEARDOWN
        \end{scope}%
  \pgfresetboundingbox
  \path[use as bounding box] (image.south west) rectangle (image.north east);
  \end{tikzpicture}%
}

\begin{document}
\title{Stochastic cellular automaton model of culture formation}
\author{Frederik Ravn Klausen}
\email{klausen@math.ku.dk}
\affiliation{QMATH, University of Copenhagen, Universitetsparken 5,
Copenhagen Ø, Denmark}
\author{Asbjørn Bækgaard Lauritsen}
\email{alaurits@ist.ac.at}
\affiliation{Institute of Science and Technology Austria, Am Campus 1, 3400 Klosterneuburg, Austria}
\date{\today}

\begin{abstract} 
We introduce a stochastic cellular automaton as a model for culture and border formation. The model can be conceptualized as a game where the expansion rate of cultures is quantified in terms of their area and perimeter in such a way that approximately geometrically round cultures get a competitive advantage.  
We first analyse the model  with periodic boundary conditions, where we study how the model can end up in a fixed state, i.e. freezes. 
Then we implement the model on the European geography with mountains and rivers. 
We see how the model reproduces some qualitative features of European culture formation, namely that rivers and mountains are more frequently borders between cultures, mountainous regions tend to have higher cultural diversity and the central European plain has less clear cultural borders.  
\end{abstract} 

\maketitle

\section{Introduction}

The topic of border formation between nations or cultures is complex and it has been subject of interdisciplinary discussion for centuries \cite[Chapter 2]{agnew2002making}. The role of natural boundaries, such as rivers and mountains, in the formation of borders was the basis of many early thoughts on borders \cite[p.22-23]{popescu2011bordering}, but has since been under criticism in the academic literature \cite{fall2010artificial}. 

Simultaneously, the statistical physics of social dynamics aims at making simple models of complex social phenomena to capture some, but not all aspects of the phenomena. This point of view has with some success been applied to areas such as traffic, networks, economics  \cite[Chapter 1]{RevModPhys.81.591}.
See also the recent collection \cite{Perc.2019} and the review \cite{Jusup.Holme.ea.2022}
for more applications of physics to the study of social phenomena.

In particular, models of culture and language inspired by statistical physics have been intensively studied in recent decades. Some of the most studied examples are the voter model \cite{holley1975ergodic}, the Axelrod model \cite{axelrod1997dissemination} along with many others (see for example the review \cite{RevModPhys.81.591}).

 In this paper, we apply the methods of statistical physics to another complex topic of the social sciences: The problem of border formation. We show that a simple model taking only the locations of seas, rivers and mountains as input can reproduce significant features of actual border locations.

The model is a stochastic cellular automaton with coarsening dynamics constructed such that approximately round cultures spread faster. It is inspired by, but yet substantially different from, the agent-based model from \cite{Dybiec.Mitarai.ea.2012} which took spreading of information as a starting. %point for their model of cultures. 
Instead, the inspiration for our model stems from ideas in popular culture about the role of geography and military power in border formation \cite{marshall2016prisoners}. For instance, the popular strategic games  \emph{Risk} \cite{risk} and \emph{Civilization} \cite{civ6} which involve using armies to conquer territory.

%Hence one should view our model merely as rough simplification of the problem that only aims at scratching the surface of the topics of borders. 
Thus, the model becomes a concrete mechanisation of 19th century naturalistic thoughts on borders where power and natural geography played a central role (although a major difference is that we present a probabilistic rather than deterministic point of view). 
%While we believe that our type of model at the expense of simplicity could be improved to capture slightly more naturalistic aspects (i.e. taking into account fertility) we emphasise that we make no claim that borders can be understood through our lens only. 

The failure of naturalistic models (such as the one considered in this paper) in describing more than just some overall probabilistic correlations may well be used as yet another argument against the 19th century naturalistic point of view that was implicitly build to fuel contemporary imperialistic agendas \cite[Chapter 3]{agnew2002making}. 
With these objections 
in mind
%on the top of our minds, 
we nevertheless construct a simple model for borders that take only natural boundaries and power into account. 
%In this paper we give a computational model for culture formation based more on power than what to our knowledge has previously been investigated. 
In particular, we ignore the myriad of other factors such as climate, diseases, individuals, cities, trade routes, taxation, technology, natural disasters, religion, crops, ideologies etc. 
Adding model features to account for these factors would come at the expense of  the simplicity of the model and we believe they may be better dealt with using other approaches. 
%We emphasise that the simplicity of our model is a feature rather than a bug.

%However, our model also entails adopting the old, arguably naive, point of view that natural boundaries, rivers, sea and mountain.  
%Similarly, focusing on the role of power in border formation is naive and doing so one will miss many important nuances in the understanding of borders. \todo{cite}

Over the last 25 years, stochastic cellular automata have been used to model similar spacial forms and reproduce dynamic spatial behaviour \cite{soares2002dinamica}. 
Examples of the approach include urban growth \cite{clarke1997self,ghosh2017application} and forest fires \cite{karafyllidis1997model}.

We study our model both with periodic boundary conditions (i.e. on a torus) with no geographical features and on a map of Europe with the geographical features of rivers and mountains. 
On the European map we compare our model to historical data of border locations in the years 1200-1790 from \cite{Abramson.2017,Carter.Ying.ea.2022,Abramson.Carter.ea.2022}.
Although we compare outcomes of our model to historical data, we emphasise that the purpose of the model is not to predict the actual borders of Europe, but rather to demonstrate a (computable, random) process capable of forming reasonable cultural borders.
Since the current European borders to some extent reflect the cultural/linguistic boundaries our model is in turn a  model of border formation. This is also the reason we focus on the European map as it could be argued that it is the region where cultural identities and political borders are the most interrelated \cite{bufon2006between}. 

%\newpage 

\section{Model: The game of Europe} 

\begin{figure}[htb]
%\floatbox[{\capbeside\thisfloatsetup{capbesideposition={left,top},capbesidewidth=4cm}}]{figure}[\FBwidth]
\centering
\includegraphics[width=\columnwidth]{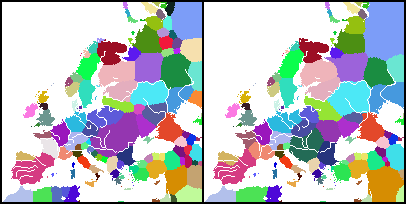} 
\caption{Snapshot of the dynamics of the model on the map of Europe at times 200 and 400.
Every country is coloured in a random colour.
Note the presence of countries of different sizes, that some rivers form clear borders and that some rivers are fully contained in one country.
} 
%\label{snapshot}
\label{fig:snapshot}
\end{figure}

We consider a $200\times 200$ pixelated map of either a torus (meaning with periodic boundary conditions) or of Europe, which consists of 40000 cells.  A \emph{country} is a set of cells and we assign each (land) cell to a country, see for example Figure \ref{fig:snapshot}. The \emph{countries} should not necessarily be thought of as real countries, but could just as well be interpreted as cultures, tribes, or even ideas.

One can envision our model as a \emph{game of Europe} where countries with varying power compete against each other using armies. We emphasise again that we only use this terminology to reflect our inspiration and for clarity of presentation.

%think about which plot to include here

At each time step, every cell is updated based on interactions between neighbouring countries. 
To define the interactions we first need to define the \emph{powers} of the different countries.
A naive approach is to define the power of a country to be its area divided by its perimeter. 
This is motivated by the following:
\begin{enumerate}
    \item The size of the army a country has is roughly proportional to its population, 
which again is roughly proportional to its area.
\item This army should be positioned in the border regions 
- having size the perimeter, as this is where the interaction with neighbouring countries takes place.
\end{enumerate}

The validity of point 1. above can well be criticised. See for instance the reference \cite{Abramson.2017}, where the size of an army is argued to be given by the wealth of the corresponding country,
or \cite{Ackland2007CulturalHO}, where fertility is considered.
The model presented here,
%Our model, 
though more naive, is much simpler, since we do not need any a priori information of wealth of different areas or some other descriptor of population, the size of an army, etc.

 If we define
%Taking 
the power of a country to be its area divided by its perimeter the power has dimensions of length, and thus large countries are stronger (more powerful). 
This leads to an uninteresting dynamics, where one country quickly dominates. 
For a more interesting dynamics we therefore
 choose
the power of a country to be some dimensionless quantity. 
Denoting the area of country $C$ by $\alpha_C$ and its perimeter by $\pi_C$ we define its power as
\begin{align} \label{eqn.def.power}
\Pi_C := \frac{\sqrt{\alpha_C}}{\pi_C}.
%\text{strength} = \frac{\sqrt{ \text{area} }}{ \text{perimeter} }.
\end{align}
There are many ways of constructing such a dimensionless power, 
e.g. %$\text{area}/\text{perimeter}^2$ 
$\alpha_C / \pi_C^2$
is another possible choice.
\Cref{eqn.def.power}
is a natural choice and leads to a more local dynamics as will be explained below.

More precisely a cell is a pair of integers $(i,j)$ and 
the area of a country is simply defined as the number of cells the country  consists of.
The perimeter is defined as the number of cells $(i,j)$ owned by the country such that
at least one of the cells in the $3\times 3$ square\footnote{For the cell $(i,j)$ the $3\times 3$ square around it means the set 
$\{i-1,i,i+1\} \times \{j-1,j,j+1\}$, i.e. the $9$ cells
$(i-1,j-1), (i-1,j), (i-1,j+1), (i,j-1), (i,j), (i,j+1), (i+1,j-1), (i+1,j), (i+1,j+1)$.
\label{note.3x3}} 
around the cell $(i,j)$ belongs to a different country.

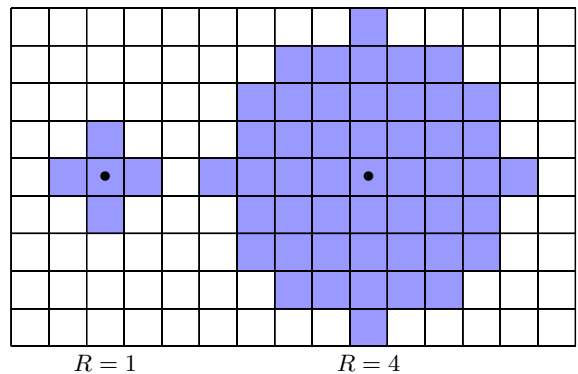
\begin{figure}[htb]
    \centering
    \begin{tikzpicture}[scale=0.5]
    \fill [blue!40] (1,4) rectangle (4,5);
    \fill [blue!40] (2,3) rectangle (3,6);
    \fill [blue!40] (5,4) rectangle (14,5);
    \fill [blue!40] (6,2) rectangle (13,7);
    \fill [blue!40] (7,1) rectangle (12,8);
    \fill [blue!40] (9,0) rectangle (10,9);
    \foreach \x in {0,1,...,15}
    \foreach \y in {0,1,...,9}
    {
    \draw (\x,0) -- (\x,9);
    \draw (0,\y) -- (15,\y);
    }
    \node at (2.5,4.5) {$\bullet$};
    \node at (9.5,4.5) {$\bullet$};
    \node[anchor=north] at (2.5,0) {$R=1$};
    \node[anchor=north] at (9.5,0) {$R=4$};
    \end{tikzpicture}
    \caption{The radius of influence determines which cells a cell influences. The blue (darker) areas are the neighbourhoods corresponding to radius of influence $R=1$ and $R=4$.     
    \label{fig:neighbour}
   }
\end{figure}

For the ``battle'' of a cell $(i,j)$ we need to define the \emph{neighbours} of $(i,j)$. All cells $(i', j')$ such that the centre of $(i', j')$ is closer to $(i,j)$ than some number $R$ are considered the neighbours of $(i,j)$. 
See Figure \ref{fig:neighbour} for an illustration. 
More formally, the relation is that $(i', j')$  is a neighbour of $(i,j)$ if  $(i'-i)^2 + (j'-j)^2 \leq R^2$. 
Notice also that a cell is a neighbour of itself.

We call $R \in \lbrack 1, \infty)$ the \emph{radius of influence}. 
Countries occupying more of the neighbouring cells are stronger, as they have more of their army in the neighbourhood. 
Denote by $N_C(i,j)$ the number of neighbours of cell $(i,j)$ of country $C$.
The local power of the country $C$ is 
its (global) power times its number of neighbourhood cells.

Finally, we add some randomness to the dynamics. This is controlled by the \emph{fluctuation} $p\in [0,1]$.
Each country's power at the cell $(i,j)$ is multiplied by independent fluctuations uniformly in $(1-p, 1+p)$.
We denote the (random) fluctuation of country $C$ in cell $(i,j)$ by $\Phi_C(i,j)$. 
(All $\Phi_C(i,j)$ are independent.)
In total, the local power of country $C$ in the battle of cell $(i,j)$ is given by
\[
    \Pi_C^{\textnormal{loc}}(i,j)
    := \Pi_C \times N_C(i,j) \times \Phi_C(i,j).
    %&\text{local-strength} = \text{strength} \times (\text{no. neighbours}) \times \text{fluctuation}.
\]
The country with the highest local power at cell $(i,j)$ wins the battle and conquers the cell $(i,j)$
(or, defends, if the country already occupied cell $(i,j)$).\footnote{One may argue that technically in order for the model to be a cellular automaton the dynamics has to be local and since the countries might be arbitrarily large the computation of power and perimeter may depend on arbitrarily many cells which contradicts locality. Since we see that our countries stay local we still call the model a cellular automaton. }

The dynamics is computed at all cells simultaneously. 
After each time step the map is updated accordingly, and the process repeats.

We can now describe how the choice of the power in \Cref{eqn.def.power} gives a more local dynamics.
The other relevant choice of the power $\Pi_C$ is the square of the choice in \Cref{eqn.def.power}. 
If a cell has $N_A$ neighbours of country $A$ and $N_B$ of country $B$
the local powers (without the fluctuation) are 
$\Pi_A N_A$ and $\Pi_B N_B$, whereas for the other choice of the power it would be $\Pi_A^2 N_A$ and $\Pi_B^2 N_B$.
For the second choice the global powers $\Pi_A$ and $\Pi_B$ matter more for the battle of some cell, while for the first (\Cref{eqn.def.power}) more weight is put on the local effect of how many neighbours $N_A$ and $N_B$ the cell has of the different competing countries.

On the map of Europe, the above-described dynamics is modified by geographic parameters. 
Further, we have to describe boundary conditions.
These geographic parameters and the boundary conditions are explained in Section \ref{sec:geo}.
On the torus (meaning the $200\times 200$ cell grid has periodic boundary conditions) there are neither geographic features nor boundary conditions and it is thereby a model with no geography.

\section{Freezing transition on the torus} \label{sec:torus}
 The model presented above
%Our model 
is a Markov chain on the set of configurations.
As such, any realisation of the dynamics will for large enough time 
end up in some irreducible recurrence class \cite[Theorem 1.40]{markov}.\footnote{We recall that a set of configurations being a recurrence class means that the dynamics will stay in the class. The set is irreducible if any configuration in the class can be reached from any other configuration, 
see \cite[Chapter 1]{markov}.}

Clearly, any configuration of only one country occupying all cells is an irreducible recurrence class (of just one configuration) and one can ask the question if these are the only ones.

To investigate this we define a configuration to be \emph{freezing} if it is a recurrence class (of just one configuration, hence in particular irreducible). 
That is, the probability of staying in a freezing configuration is $1$ and as such the dynamics ``freezes'' if it reaches such a configuration.
In particular, the one-country configurations are freezing.

For small fluctuations $p$, there may be other freezing configurations, but for $p$ large there are none.
Even, for large enough $p$ ($p=1$ is trivially large enough) any other configuration is transient, meaning that, with probability $1$, the dynamics eventually ends up in a one-country configuration.
Thus there exists a smallest $p$, the critical $p_c$, such that for any $p>p_c$ the only irreducible recurrence classes are the one-country configurations.

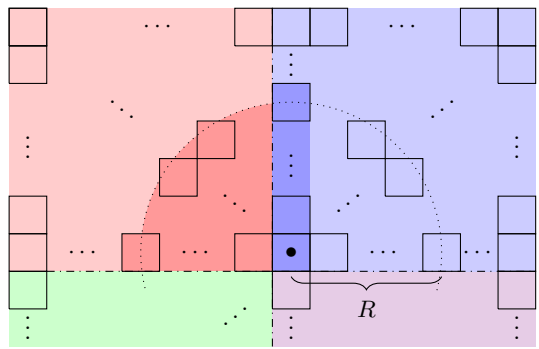
\begin{figure}[htb]
    \centering
    \begin{tikzpicture}[scale=0.5]
        %\foreach \x in {-4,-5,...,4}
        %\foreach \y in {0,1,...,4}
        %{
        %    \node (\x,\y) at (\x, \y) {};
        %}
        %\foreach \x in {-4,-3,...,4}
        %\foreach \y in {0,1,...,4}
        %{\draw[fill] (\x,\y) circle [radius=1.5pt];}    
        \fill [red!20] (-7.5,-0.5) rectangle (-0.5,6.5);
        \fill [green!20] (-7.5,-2.5) rectangle (-0.5,-0.5);
        \fill [violet!20] (-0.5,-2.5) rectangle (6.5,-0.5);
        %\fill [red!40] (0,0) -- (0,5) arc[start angle=90, end angle=180,radius=5] -- (0,0);
        \fill [blue!20] (-0.5,-0.5) rectangle (6.5,6.5);
        %\fill [blue!20] (0.5,4.5) rectangle (1.5,5.5);
        \begin{scope}
        \clip (4,0) arc[start angle=0, end angle=360,radius=4];
        \fill[red!40] (-5,-0.5) rectangle (-0.5,5);
        \end{scope}
        \foreach \x/\y in {-2/3,-3/2,-4/0}
        {\fill [red!40] (\x-0.5,\y-0.5) rectangle (\x+0.5,\y+0.5);}
        \fill [blue!40] (-0.5,-0.5) rectangle (0.5, 4.5);
        \begin{scope}
        \clip (-6,-1) rectangle (6,6);
        \draw[dotted] (4,0) arc[start angle=0, end angle=360,radius=4];
        \end{scope}
        \draw[dashdotted] (-6.5,-0.5) -- (6.5,-0.5);
        \draw[dashdotted] (-0.5,-2.5) -- (-0.5,6.5);
        \foreach \x/\y in {-7/0,-4/0,-1/0,0/0,1/0,4/0,0/1,0/4,-3/2,-2/3,2/3,3/2,-7/1,6/0,6/1,-7/5,-7/6,-7/6,5/6,6/5,6/6,-1/6,0/6,1/6,-7/-1,0/-1,6/-1}
        {
        \draw (\x-0.5,\y-0.5) -- (\x+0.5,\y-0.5) -- (\x+0.5,\y+0.5) -- (\x-0.5,\y+0.5) -- (\x-0.5,\y-0.5);
        }
        \foreach \x/\y in {-2.5/0,2.5/0,3/6,-3.5/6,-5.5/0,5/0}
        {
        \node (\x\y) at (\x,\y) {$\cdots$};
        }
        \foreach \x/\y in {-1.5/1.5,-4.5/4}
        {
        \node (\x\y) at (\x,\y) {$\ddots$};
        }
        \foreach \x/\y in {1.5/1.5,4/4,-1.5/-1.5}
        {
        \node (\x\y) at (\x,\y) {$\iddots$};
        }
        \foreach \x/\y in {0/2.5,-7/3,6/3,0/5.2,-7/-1.8,0/-1.8,6/-1.8}
        {
        \node (\x\y) at (\x,\y) {$\vdots$};
        }
        \draw[decoration={brace,mirror,raise=2pt,amplitude=5pt},decorate] (0,-0.5) -- node[below=8pt] {$R$} (4,-0.5);
        %\draw[decoration={brace,mirror,raise=2pt,amplitude=5pt},decorate] (6.5,0) -- node[right=8pt] {$k$} (6.5,6);
        \node (cor) at (0,0) {$\bullet$}; 
        %circle [radius=2pt];
    \end{tikzpicture}
    \caption{Schematic of the battle of a corner cell ($\bullet$) in a {tiling configuration} with the defending country (top right) in light blue and a strongest attacking country (top left) in light red.
    %and attacking in red. 
    The dark red shaded region (left darker region) is the attacking (red) cells in the neighbourhood. Note that the neighbourhood (see also \Cref{fig:neighbour}) can be decomposed into the origin (meaning the corner cell) and $4$ regions congruent to the dark red shaded region (left darker region). Hence $N_R = 4N_A + 1$, with $N_R$ denoting the number of cells in a neighbourhood of radius $R$ and $N_A$ denoting the number of attacking cells of a strongest attacking country.
    The defending country has exactly $R+1$ cells more than a strongest attacking one (say in the central column (in dark blue)) so $N_D = N_A + R +1 = \frac{N_R}{4} + R + \frac{3}{4}.$
    \label{fig:freeze}}
\end{figure}

A necessary condition for the only irreducible recurrence classes being the one-country configurations is that no other configuration is freezing.
We calculate here the minimal value $p_{\textnormal{tile}}$ for 
a \emph{tiling configuration}  (meaning all countries are $k\times k$ squares for a $k > 2R$, see \Cref{fig:freeze}) 
to be \emph{non-freezing}. 
Trivially then $p_c \geq p_{\textnormal{tile}}$.

%However on the infinite lattice $\Z^2$ and for radius of influence $R=1$
%we have $p_c = p_{\textnormal{tile}} = \frac{1}{2}$ as we show in \Cref{sec:analytic_freezing} below.
%We do not know whether $p_c = p_{\textnormal{tile}}$ in general on $\Z^2$. 

It is a simple calculation to see that for a 
{tiling configuration} the cells that are easiest to conquer are the corners. All countries have the same power (they all have the same shape) and thus we need
\begin{equation}\label{eqn.freeze.ini}
    (1-p)N_D < (1+p)N_A
\end{equation}
for such a tiling configuration to be non-freezing. 
Here $N_D$ ($N_A$) denotes the number of neighbours of the corner of the defending (respectively strongest attacking) country 
and we suppress in the notation the dependence on the cell $(i,j)$.
A counting argument (see \Cref{fig:freeze}) finds that 
$N_D = \frac{N_R}{4} + R + \frac{3}{4}$ and 
$N_A = \frac{N_R}{4} - \frac{1}{4}$
for integer $R$,
where $N_R$ denotes the number of cells in the neighbourhood corresponding to radius of influence $R$ (see \Cref{fig:neighbour}).
Thus, \Cref{eqn.freeze.ini} gives 
\[
p_{\textnormal{tile}} = \frac{2(R+1)}{N_R + 2R+1}.
\]
For large $R$ we have $N_R \simeq \pi R^2$ and so 
$p_{\textnormal{tile}}\simeq 2/\pi R$. 
For small $R$ the values of $p_{\textnormal{tile}}$ are given in \Cref{table:freeze}.
\begin{table}[htb]
    \centering
    \caption{Minimal value $p_{\textnormal{tile}}$ for a tiling configuration to be non-freezing. 
    We see that $p_\textnormal{tile}$ decreases in $R$ meaning that for larger $R$ less fluctuation is needed for the tiling configuration to be non-freezing.}
    %The table highlights the relationship between $p$ and $R$ 
    %that entails that when the radius increases the dynamics become faster. }
    \label{table:freeze}
    \begin{tabular}{|c||c|c|c|c|c|c|c|c|}
    \hline
        $R$ & 1 & 2 & 3 & 4 & 5 & 6 & 7 & 8 \\
        \hline
        $p_{\textnormal{tile}}$ 
        & 0.500
        & 0.333
        & 0.222
        & 0.172
        & 0.130
        & 0.111
        & 0.0976
        & 0.0841
        \\
        \hline
    \end{tabular}
\end{table}

Finally, an empirical observation is that for large $R$ ($R=6,7,8$) the shapes of the countries in the simulations turn out to be approximately hexagonal after many time steps. A configuration of hexagonally shaped countries is however non-freezing for $p$ smaller than $p_{\textnormal{tile}}$, but it does signify a metastable configuration.
For the value $R=4$ used in most of the simulations however, we do not see such hexagonal shapes. Moreover, countries of different sizes coexist, (as we also see on the European map in \Cref{fig:snapshot}).
%\todo[inline]{bedre formulering? / Eller slet? Jeg har forsøgt en bedre formulering}

\section{Geographical features}\label{sec:geo}
We now describe the geographical features of Europe taken into account in the model. 
This differs from the work \cite{Dybiec.Mitarai.ea.2012}, where only the coastline is incorporated.
The parameters of the model are summarised in \Cref{table:1}.

\paragraph*{Coasts and boundary conditions:} 
We implement the sea as cells that cannot be conquered. 
Further, if a cell is on the perimeter of a country only because it borders the sea 
(meaning its $3\times 3$ square (see \Cref{note.3x3} on page \pageref{note.3x3}) contains only sea and its own country) 
we add to the perimeter the sea-border parameter $s \in \lbrack 0,1 \rbrack$ instead of $1$.

The parameter $s$ models how much easier it is to defend a country which has the sea instead of neighbours. In all of the following we set $s=0.5$ we therefore do not include it as a parameter. 
Similarly, for the boundaries of the map we choose the same parameter $s$ - our main goal with penalising the boundaries (which are beyond the Ural, Kaukasus and Sahara) is to make sure that the boundary conditions do not influence the centre of the map.

\paragraph*{Rivers} 
are like the sea implemented as cells that cannot be conquered
and give the same effect to the perimeter as the sea, through the parameter $s$ (that we set to $0.5$).
Additionally, if a country occupies both sides of a river, it gets a bonus to its area through the parameter 
$A_r \geq 0$ and the river no longer counts as a border. 
Technically this is implemented as follows. Suppose cell $(i,j)$ is part of the river. 
If all cells in the $3\times 3$ square around cell $(i,j)$ (see \Cref{note.3x3} on page \pageref{note.3x3}) are either part of the river, 
or belong to the same country, then that country gets a bonus to its area of $A_r$.

\begin{figure}[htb]
\includegraphics[width=\columnwidth]{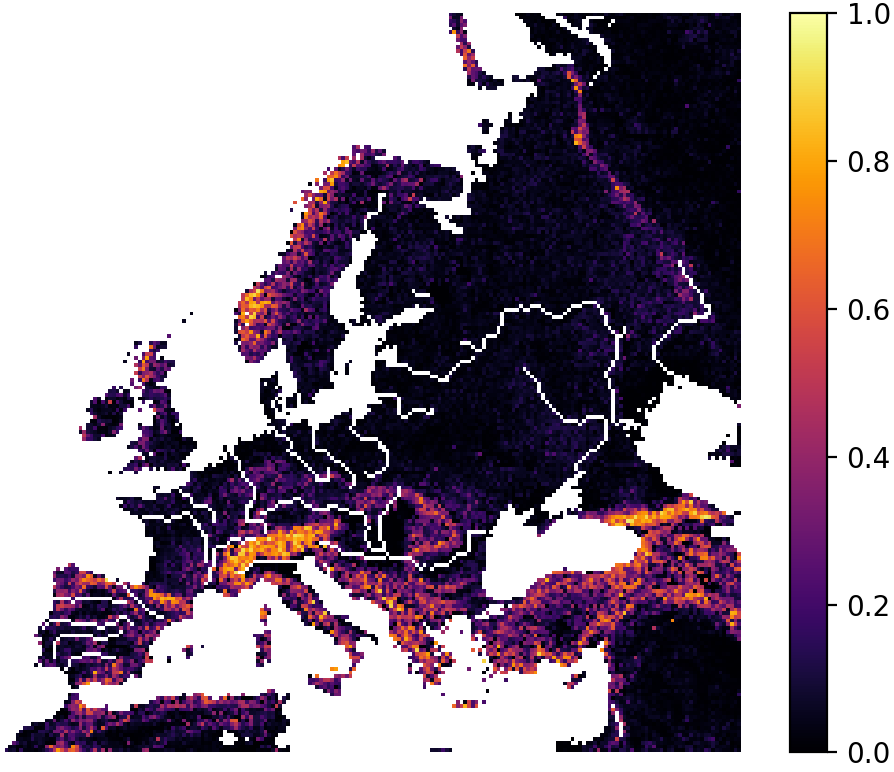}
\caption{Map of Europe with a plot of the value $m$ (colour gradient) of how mountainous the region is together with the rivers included (white inland cells).   
\label{geo}}
\end{figure}

The parameter $A_r$ is supposed to capture the effect that rivers work as trade routes as well as giving benefits to fertility and infrastructure, as many civilisations arose around rivers \cite[p.1]{mauch2008rivers}.
The location of the implemented rivers is seen in \Cref{geo}.

\paragraph*{Mountains}are implemented by a variable $m(i,j)\in[0,1]$ at each cell $(i,j)$ indicating how mountainous the cell is. 
Mountains have two effects. 
The first effect is that 
when calculating the local powers at cell $(i,j)$ the country already owning the cell gets a 
bonus parametrised by the parameter $D_m > 1$,
i.e. for the defending country $D$ its local power is
\[
    \tilde\Pi^{\textnormal{loc}}_D(i,j)
    = \left[(1-m(i,j)) + D_m m(i,j)\right] \times 
    \Pi_D^{\textnormal{loc}}(i,j).
    %\text{local-strength} = ((1-m) + D_m m) \times \text{old-local-strength}.
\]
Thus a ``fully mountainous'' region with $m(i,j)=1$ gets the defensive bonus $D_m$.

The second effect is that the perimeter contribution of cell $(i,j)$ is weighted by 
\[
    \pi_{\textnormal{w}}(i,j) = (1-m(i,j)) + P_m m(i,j)
    %\text{perimeter-weight} = (1-m) + P_m m,
\]
where $P_m \in [0,1]$ is a parameter. Here a ``fully mountainous'' cell with $m(i,j)=1$ contributes 
only $P_m$ to the perimeter and a ``completely flat'' cell  with $m(i,j)=0$ contributes 
$1$ to the perimeter.
For a country $C$ a cell $(i,j)$ in $C$ is in the perimeter of $C$ if at least one of the cells in the $3\times 3$ square around the cell $(i,j)$ belongs to a different country (see \Cref{note.3x3} on page \pageref{note.3x3}). 
The total perimeter $\pi_C$ is then calculated as the sum of $\pi_{\textnormal{w}}(i,j)$ for all cells in the perimeter of $C$.

The parameters $D_m, P_m$ model how much easier it is to defend mountainous territory. 
The parameter $D_m$ models a local effect - it is easier to defend, 
while $P_m$ models a global effect - the army of the country may be stationed elsewhere. 
The values of $m$ are plotted in \Cref{geo}.

\begin{table}
\begin{center}
\caption{Table of the parameters of the model, their symbols, and their possible/standard values.
The standard values are discussed in Section \ref{sec.results}.}
\label{table:1}
\begin{tabular}{||c c c c ||} 
 \hline
Parameter & symbol  & values  & standard choice \\ [0.5ex] 
 \hline\hline
 \multicolumn{4}{||c||}{\textbf{Base parameters}}
 \\
 \hline
 Fluctuation  & $p$  &  $\lbrack 0, 1\rbrack$  & 0.2 \\ 
 \hline
  Radius  & $R$  &  $\lbrack 1, \infty )$  & 4 \\ 
 \hline
% Border weight & $s$ &   \lbrack 0, 1\rbrack  \\
 \hline
 \multicolumn{4}{||c||}{\textbf{Geographical parameters}}  \\ 
 \hline
 River area & $A_r$ &   $\lbrack 0, \infty)$ &  8 \\
 \hline
Mountain defence & $D_m$ &  $\lbrack 1, \infty)$ &  2 \\
 \hline
Mountain perimeter & $P_m$ &  $\lbrack 0, 1\rbrack$ & 0.5 \\ %[1ex] 
 \hline
\end{tabular}
%\vspace*{-1em}
\end{center}
\end{table}

\section{Methods} \label{methods} 
In this section, we briefly describe the sources of data and how we prepare the data.  
The mountainous parameter is defined via the 
elevation data from the GMTED2010 dataset \cite{elevation.data,gmted.2011}.
Major rivers in the EU are from the WISE Large rivers and large lakes dataset \cite{rivers1} 
and rivers Don, Volga, and Ural are from \cite{Basher.Lynch.ea.2018}.
All data is changed to the Lambert Azimutal Equal Area projection using QGIS \cite{QGIS}.
From the elevation data, the value $m$ is calculated as follows.
For each cell, the mean deviation from its height (in meters) to the other 8 cells in the $3\times 3$ square around it (see \Cref{note.3x3} on page \pageref{note.3x3})
is computed as the ``curvature'' $\kappa$ (in meters), then the mountainous parameter $m$ is calculated as 
$m = 1-\exp(-\kappa/100)$. The exact function used here is not that important.
Essentially, we just need that $m$ interpolates between flat regions where $m\simeq 0$ and mountainous regions where $m \simeq 1$. 
Finally, the resolution of the data has been reduced to the $200\times 200$ grid used. The rivers have been hand-curated to fit the $200 \times 200$ grid reasonably. The total number of land cells totals $20787$. 
The result can be seen in Figure \ref{geo}. 
The code and datasets are available at Github \cite{Klausen_model-for-culture-formation_2023}
and ISTA \cite{Klausen.Lauritsen.2023}.\footnote{The simulations were conducted on a 2010 laptop with a dual-core Intel Core i5-5257U processor and 8 GB RAM for about one year.}

The data of historical borders are from \cite{Carter.Ying.ea.2022,Abramson.Carter.ea.2022}, which in turn is based on \cite{Abramson.2017}.
The data consists of maps of Europe divided into states at 5 year intervals in the period 1200-1790 (i.e. the years 1200, 1205, 1210 and so on). 
What exactly constitutes a ``state'' is discussed in \cite{Abramson.2017}, where also the dataset is described in more detail.
The dataset does not cover the full extent of the map we use, in particular for the earlier years. 
It does however cover most of Europe (apart from the earlier years, where northern Norway, Sweden and Finland and Russia east and immediately west of the Urals is not covered),
see \Cref{fig:Historic}.
The maps of \cite{Carter.Ying.ea.2022,Abramson.Carter.ea.2022,Abramson.2017}
are changed to the Lambert Azimuthal Equal Area projection using QGIS \cite{QGIS}
and rasterized to the desired resolution using the python package ``geocube'' \cite{geocube}.
By lowering the resolution of the data to the desired $200\times 200$ grid, 
some of the smaller countries disappear, as their area is smaller than one cell.

%To gauge the effect of the parameters $p,R$ we do simulations on both the torus and map of Europe for various values of $p$ and $R$. As a starting point for the investigations of the parameters we choose $(p,R) = (0.2, 4)$, which were found to give reasonable dynamics on the European map. Interestingly, the choice $(p,R) = (0.2, 4)$ is just above the parameters of freezing for the torus (see Section \ref{sec:torus}).

\section{Results}
\label{sec.results}

\paragraph*{Simulations on the torus:}
To gauge the effect of the parameters $p,R$ on the torus we 
perform
simulations for various values of $p$ and $R$.
As a starting point for the investigations of the parameters we choose $(p,R) = (0.2, 4)$, which were found to give reasonable dynamics on the European map. Interestingly, the choice $(p,R) = (0.2, 4)$ is just above the parameters of freezing for the torus (see Section \ref{sec:torus}).

\begin{figure}[b]
%\begin{figure*}

\centering
%\begin{captivy}{
\begin{tabularx}{\linewidth}{*{3}{X}}
\includegraphics[width=\linewidth]{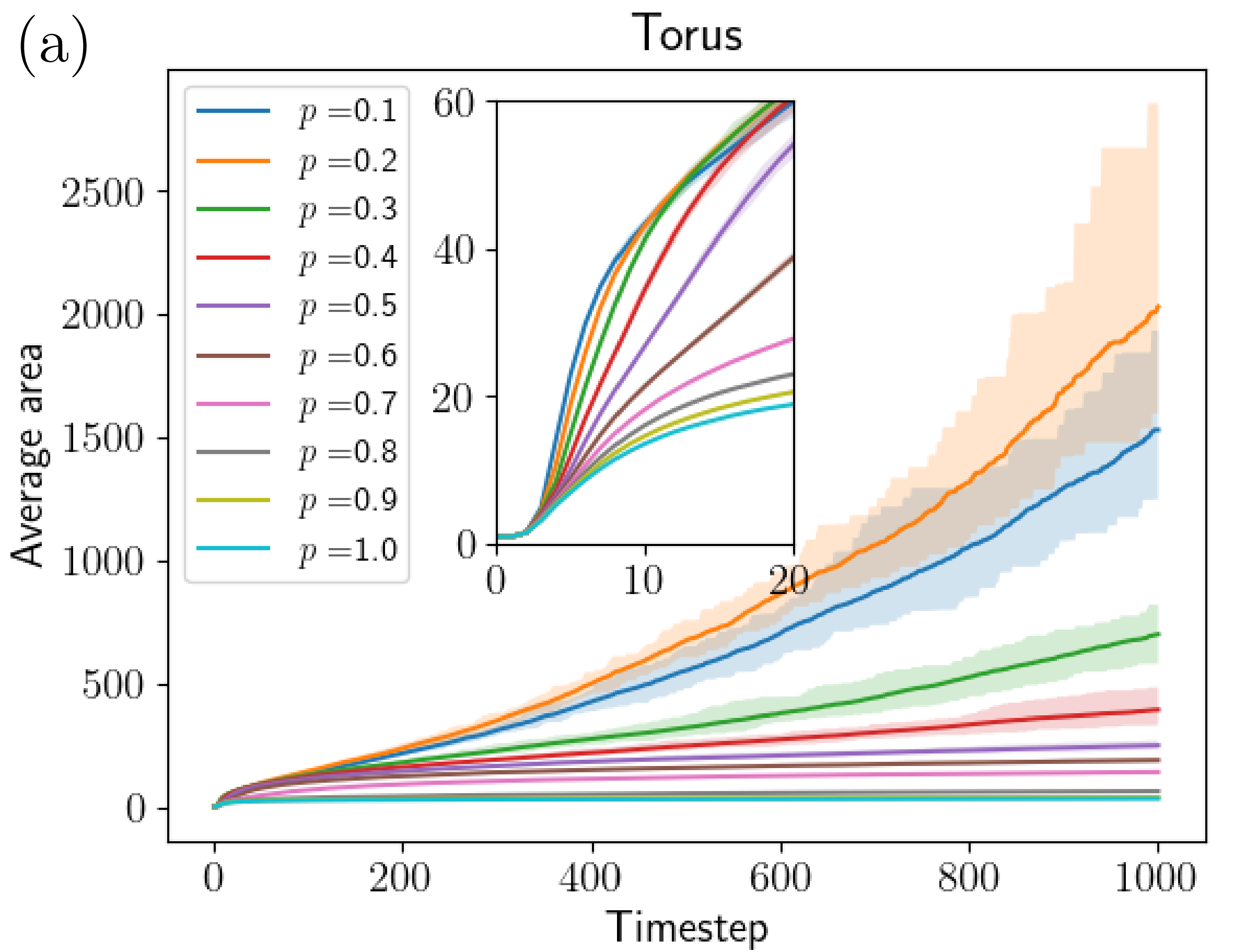}
&
\includegraphics[width=\linewidth]{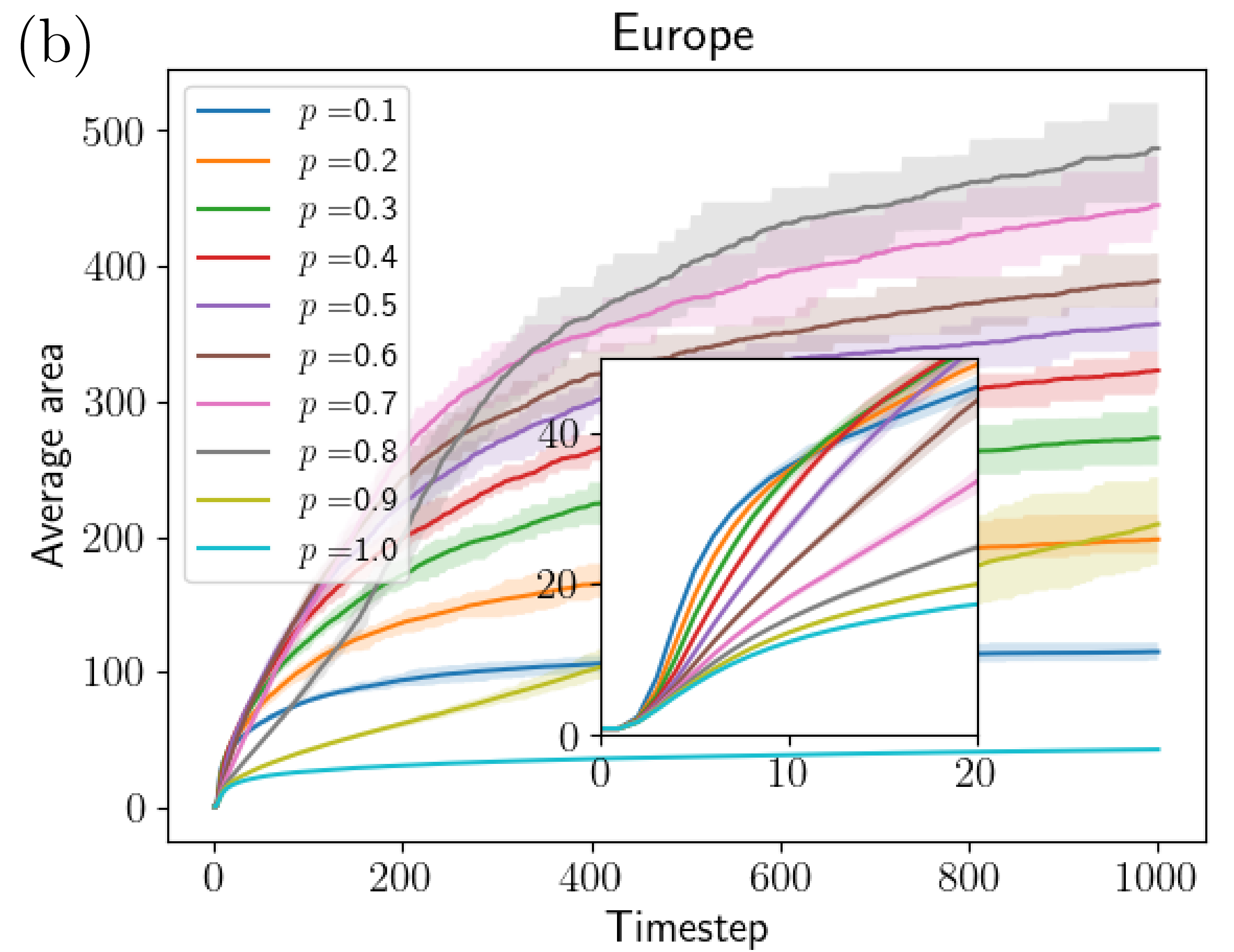}
&
\includegraphics[width=\linewidth]{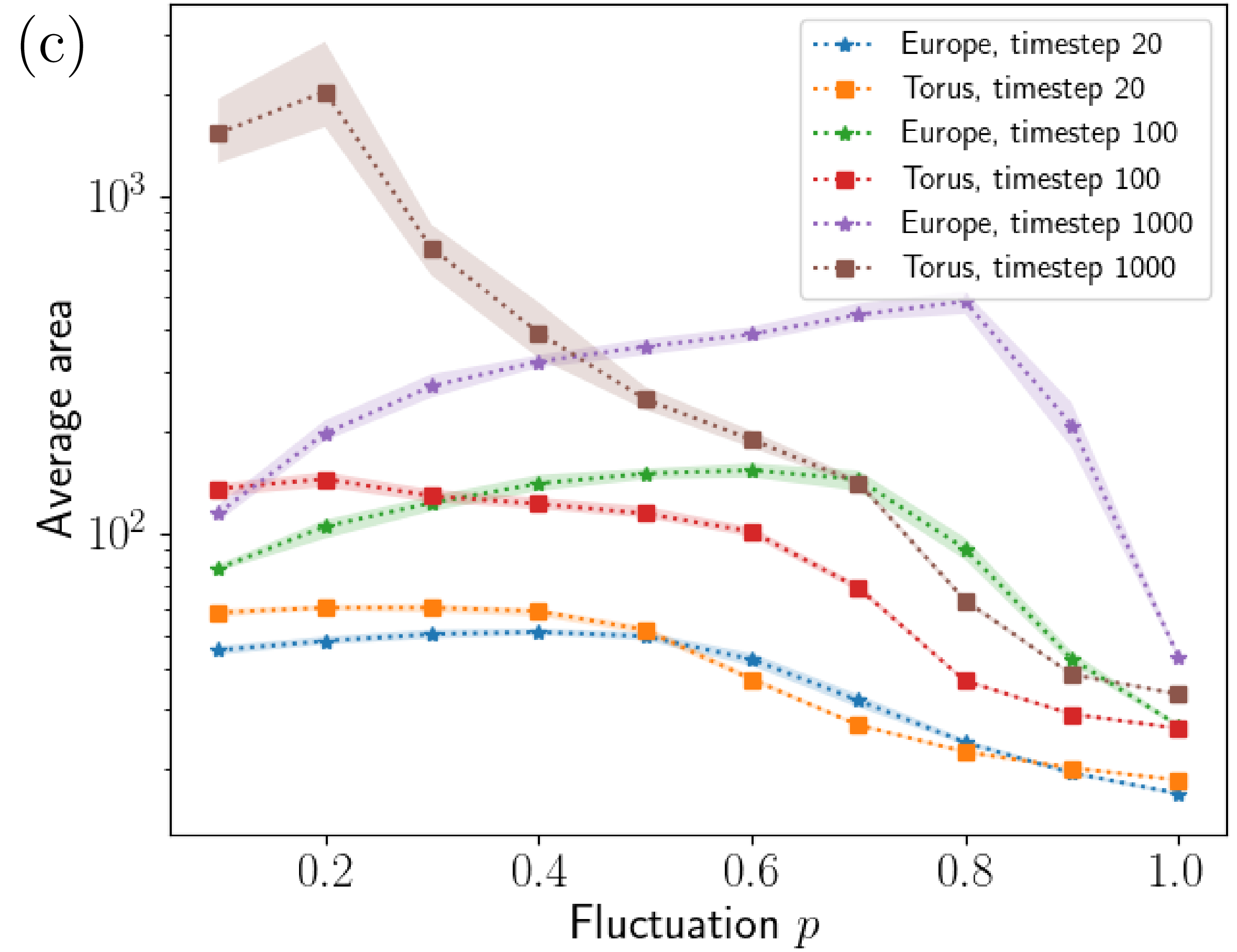}
\end{tabularx}
%}
%\oversubcaption{0.02, 0.95}{}{fig:area_torus_sub}
%\oversubcaption{0.35, 0.95}{}{fig:area_europe_sub}
%\oversubcaption{0.70, 0.95}{}{fig:area_fluc_sub}
%\end{captivy}
\caption{((a) and (b))  Averaged areas of countries over time for different fluctuations $p$ for simulations on the torus (a) and the map of Europe (b).
For the simulations on the torus the parameters are chosen to be 
$R = 4$ and for the map of Europe $A_r=8, D_m=2, P_m=0.5, R=4$.
(c)
Plot of the average areas as a function of the fluctuation parameter $p$ for both simulations on the torus and on the map of Europe at time step 20,100 and 1000.
All plotted data are averages over 20 simulations and the shaded regions are between the $5\%$ and $95\%$ quantiles.
Notice the very different behaviour of the dynamics on the torus and the map of Europe (for large times) with the fastest evolution taking place for very different values of $p$. 
In (c) the fastest evolution corresponds to maximum of the curves shown. 
In the inserts of (a) and (b) we see that for very short times, the evolutions on the torus and map of Europe are very similar. 
\vspace{1em}
}
\label{fig:areas_p_torus}
\label{fig:areas_p_Europe}
\label{fig:p_vs_area}

%\end{figure*}
\end{figure}

\paragraph*{Fluctuations:} 
In 
%\Cref{fig:areas_p_torus,fig:p_vs_area} (\subref{fig:area_torus_sub} and \subref{fig:area_fluc_sub})
Figures \ref{fig:p_vs_area}(a) and \ref{fig:p_vs_area}(c) 
%\Cref{fig:area_torus_sub,fig:area_fluc_sub} 
we vary the fluctuations and plot the average country size as a function of time respectively the fluctuation. 
In \Cref{fig:p_vs_area}(c) we see that at time step number $1000$ the average country size has a maximum (in $p$) around $p=0.2$. 
In particular, the model is non-monotone in the fluctuation $p$ 
and as such, $p$ cannot be interpreted as an effective temperature of the model.

\paragraph*{Radius:}
For completeness, we also study the effect of the radius.
These findings are shown in \Cref{fig:areas_R_torus} in the Appendix. We see that the average size of countries is monotone increasing in the radius of influence $R$.

\paragraph*{Simulations on the map of Europe:}
In this section, we vary the parameters to determine their effects in the European geography. As a starting point, we again choose $(p,R) = (0.2, 4)$.

Fixing $R=4$ the value $p=0.2$ is close to the value of fastest growth on the torus. On the European map, however, the choice $(p,R) = (0.2, 4)$ is very slow evolving and barely above frozen (see %\Cref{fig:area_europe_sub}
\Cref{fig:p_vs_area}(b)). 
As for the geographical parameters $(A_r, D_m, P_m)$ we investigate their effects in the Appendix and find that reasonable choices are $(A_r, D_m, P_m) = (8, 2, 0.5)$.
This leads us to the standard choice of values $(p,R, A_r, D_m, P_m) = (0.2, 4, 8, 2, 0.5)$ reported in Table \ref{table:1} that we use for most of our investigations.

\paragraph*{Fluctuations:} 
In Figures 
\ref{fig:p_vs_area}(b) and \ref{fig:p_vs_area}(c)
%\Cref{fig:area_europe_sub,fig:area_fluc_sub} 
we vary the fluctuation size and plot the average country area as a function of time respectively the fluctuation. 
As for the torus we see a non-monotone dependence of the average country size in the fluctuation $p$.
For small fluctuation $p$ (i.e. below $\simeq 0.8$) larger fluctuation leads to a larger average size at time step $1000$, 
but for large fluctuation sizes the dynamics reverses.

Interestingly, the fluctuation $p$ giving the maximal growth is quite different for the European map compared to that of the torus. This may be understood as follows.

On the map of Europe, some small island and peninsula states exist forcing the average country size to be small, even if mainland Europe is split between few large countries.
This effect explains why (for large times) the average country size on the map of Europe is much smaller than that on the torus 
(see 
%\Cref{fig:area_fluc_sub}
\Cref{fig:p_vs_area}(c)).
This additionally explains the difference in which $p$'s give the largest average sizes. Namely, on the European map, the average country size is essentially given by the reciprocal of the number of such island and peninsula states. The dependence of the number of such  states on the fluctuation $p$ is a completely different dynamics
than that of the number of countries on the torus.

\paragraph*{Geographical parameters:} 
In the Appendix,  we discuss the effects of the parameters $A_r, P_m$ and $D_m$ and show how the parameters $P_m$ and $D_m$ contribute in two very different ways. 

Noticeably the parameter $D_m$ is the most influential. The effect of changing the parameter $P_m$ in comparison is much smaller.
Finally, the average country size is monotone increasing in the parameter $A_r$.

\onecolumngrid

%\clearpage

\twocolumngrid

\subsection{Findings}
We finally present the main findings. Namely, that the model, with  appropriate parameters, reproduces some of the qualitative features of the cultural borders in Europe.
In particular, we find that mountainous regions have a higher frequency of borders, which can be interpreted as higher cultural diversity in mountains. This effect is also present in historical data.

\begin{figure}[htb]
%\floatbox[{\capbeside\thisfloatsetup{capbesideposition={left,top},capbesidewidth=4cm}}]{figure}[\FBwidth]
{\includegraphics[width=\columnwidth]{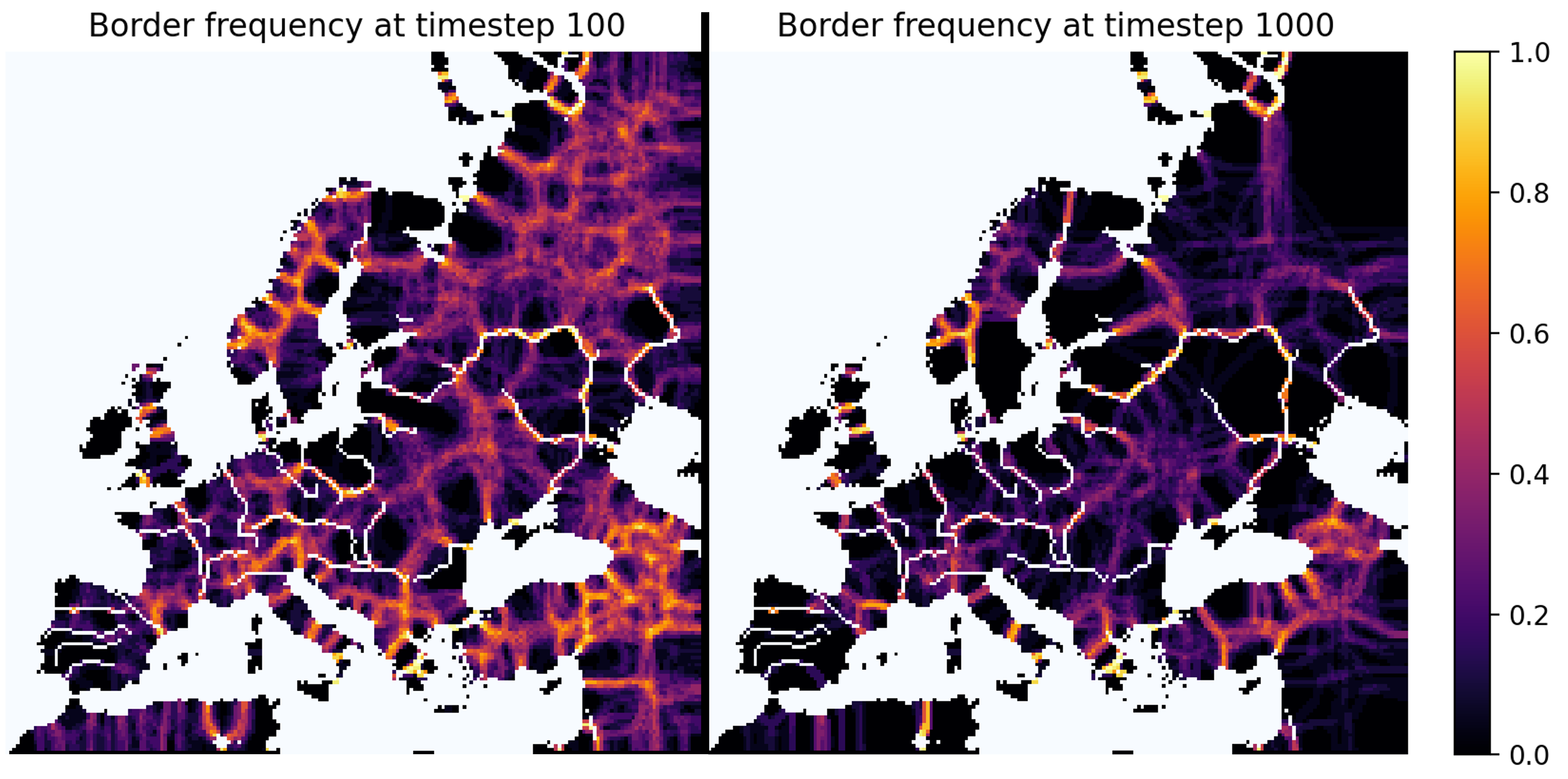}}
{\caption{Heatmap of the frequencies of when cells are borders (colour gradient)
at the time steps $t=100$ (left) and $t=1000$ (right). The parameters are chosen to be 
$p= 0.2, A_r = 8, D_m = 2, P_m = 0.5, R = 4$ and the frequencies are computed over an average of 20 iterations. 
We see that the border frequencies around rivers are significantly increased.
Simultaneously, the figure shows the coarsening effect of the model from timestep 100 to timestep 1000. Furthermore, comparing to Figure \ref{geo} we see that mountain areas tend to have higher border frequency.  \label{fig:heat_map}}}
\end{figure}

\begin{figure}[htb]
\centering
%\floatbox[{\capbeside\thisfloatsetup{capbesideposition={left,top},capbesidewidth=4cm}}]{figure}[\FBwidth]
\includegraphics[width=\columnwidth]{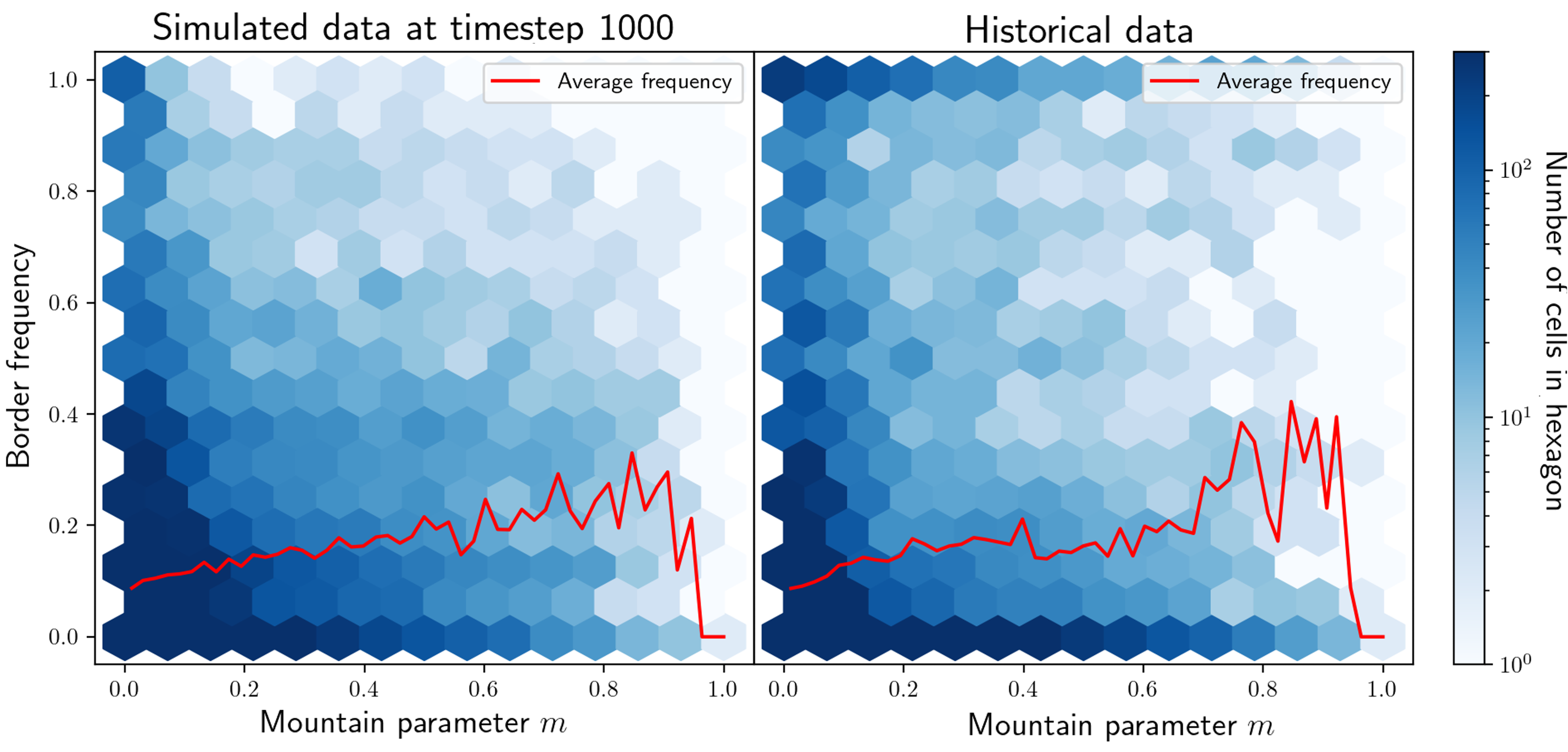}
\caption{(Left) A binned plot of border frequencies compared to the mountain parameter $m$ at  time step $t=1000$. The parameters are chosen to be 
$p= 0.2, A_r = 8, D_m = 2, P_m = 0.5, R = 4$ and the frequencies are computed over an average of 20 iterations. The (red) line is the average border frequency for bins of points with similar mountain parameter. 
 The colour gradient denotes the number of cells falling inside each hexagonal region of the plot.
(Right) A binned plot of historical border frequencies according to the mountainous parameter $m$. 
The historical data is the average of the data from \cite{Abramson.2017,Abramson.Carter.ea.2022,Carter.Ying.ea.2022}. We see that the trendline of the historical and simulated match closely. However, looking at the bins one can see that the historical data have many cells with border frequency 1 as opposed to the simulated data. One reason for this difference is the correlated nature of the historical data.}
\label{fig:borders_mountain_simulated}
\end{figure}

\paragraph*{Locations of borders:} 
In the following, we say that a land cell corresponding to one country is a \emph{border} if at least one cell in its $3\times 3$ square (see \Cref{note.3x3} on page \pageref{note.3x3})
belongs to a different country. 
In \Cref{fig:heat_map} the frequency for each cell to be a border is shown and 
in \Cref{fig:borders_mountain_simulated} we plot the correlation of the border frequency and the mountainous parameter $m$.

We see that mountain regions are more frequently borders and that larger areas that are flatter tend to have a very low density of borders. In this way the model reproduces the idea of mountains acting as natural borders, which we can also confirm in historical data in Figure \ref{fig:borders_mountain_simulated} (see also the discussion below). Further, this reflects the higher cultural diversity that is often seen in mountainous areas
\cite{debarbieux2012mountain}.

Inspecting Figure \ref{fig:heat_map} one sees that the frequency of borders close to rivers are significantly increased (compare also with the snapshot Figure \ref{fig:snapshot}), an effect observed in current subnational borders in \cite{Popelka.Smith.2020}.

\paragraph*{Historical comparison:}
%\emph{Historical comparison:} 
We next compare to the historical data from \cite{Abramson.2017,Abramson.Carter.ea.2022,Carter.Ying.ea.2022}. 
In \Cref{fig:borders_mountain_simulated} we plot the correlation between border frequencies and the mountainous parameter $m$ also for the historical data from \cite{Abramson.2017,Abramson.Carter.ea.2022,Carter.Ying.ea.2022}.
The historical borders are of course correlated since the different data points of \cite{Abramson.2017,Abramson.Carter.ea.2022,Carter.Ying.ea.2022} are only separated by 5 year intervals. These correlations, in the form of static borders, are the cause of the border frequency being 1 for many cells with low mountain parameters. 
However, there is no inherent reason to use one year over another. We thus plot  in \Cref{fig:borders_mountain_simulated} the average border frequency of the historical data and compare the simulated data to these.
As with the simulated data, we see a correlation between border frequencies and how mountainous an area is:
More mountainous regions are more often borders.
The verifies historically the idea of mountains acting as natural borders as we also see for our simulated data.

\begin{figure}[htb]
\includegraphics[width=0.8\columnwidth]{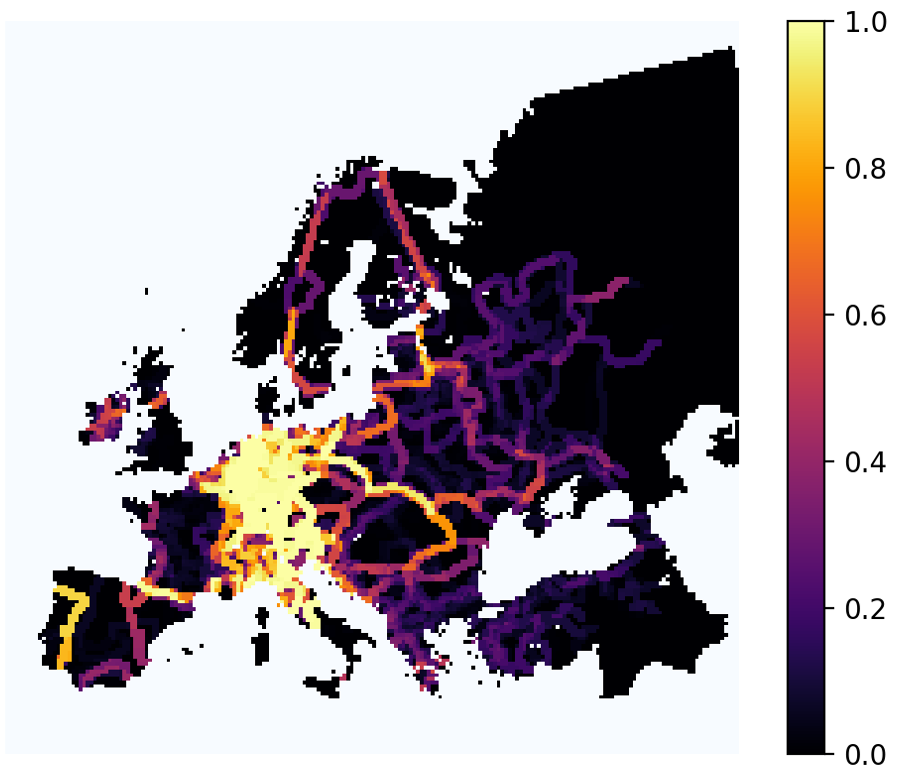}
\caption{The proportions of the time between 1200 and 1790 that each cell was a border region 
 (colour gradient) using averaged versions of the historical maps from \cite{Abramson.Carter.ea.2022,Carter.Ying.ea.2022,Abramson.2017} discussed in Section \ref{methods}. 
In particular, the dataset from \cite{Abramson.Carter.ea.2022,Carter.Ying.ea.2022,Abramson.2017}
does not define the Holy Roman Empire as a country. That leads to a high density of historical borders in central Europe.} \label{fig:Historic}
\end{figure}

In Figure \ref{fig:Historic} we plot the border frequencies of the historical data from \cite{Abramson.2017,Carter.Ying.ea.2022,Abramson.Carter.ea.2022}.
We see that the area that was then the Holy Roman Empire has a very high density of borders. This is because the Holy Roman Empire was not considered a country in the source of the data \cite{Abramson.Carter.ea.2022,Carter.Ying.ea.2022,Abramson.2017}. Instead, all the smaller individual states, usually German, that were part of the Holy Roman Empire were considered (a phenomenon sometimes known as Kleinstaaterei \cite{whaley2011germany}). 
This digression illustrates the point that our model is not a model for borders between territorial states, 
 nor for the concrete historical borders of Europe,
but rather for the cultures that might to some extend predate the emergence of the territorial state.
In particular, one should not compare the specific border frequencies in \Cref{fig:heat_map} from the model with the historical data in \Cref{fig:Historic}.

\paragraph*{Contested areas:} 
\begin{figure}[htb]
%\vspace{1em}
\includegraphics[width=0.8\columnwidth]{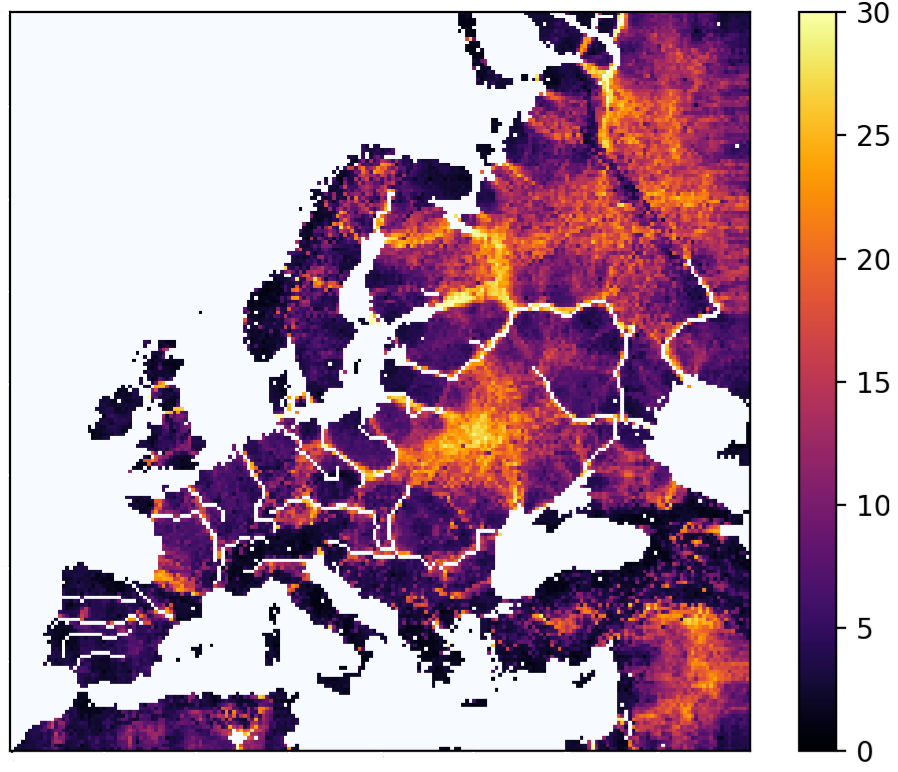}
\caption{Heatmap of the average number of times each cell has been conquered before time step $1000$ (colour gradient). The scale is cut-off at 30 as some squares have been conquered much more than 30 times. The parameters are chosen to be 
$p= 0.2, A_r = 8, D_m = 2, P_m = 0.5$ and the result is averaged over 20 simulations.
Note that highly mountainous regions like the Alps and the Caucasus are very low on the scale whereas the central Eastern European plain tops the scale.  
} 
\label{fig:contested}
\end{figure}
Finally, we consider which cells are most contested.
For each square we find how often it was conquered. This is then averaged over 20 simulations. The results are shown in \Cref{fig:contested}. 
We see that the average number of times a cell changed countries is higher on the central European plain than elsewhere, in particular in the mountainous regions in the Alps and Caucasus. 
This might be connected to the, from one point of view, less clear cultural borders on the central European plain (as can be seen on old historical maps \cite{map1, map2}, although one should be wary of context of such maps).    
Together with the above discussion on border location we reach the conclusion that not only are mountain regions more often borders, they are also much more stable borders.

\section{Discussion and outlook}
We have seen that for suitable choices of the parameters, the model efficiently reproduces many of the features of historical political and cultural borders:

There is a high density of borders in mountain regions and along rivers and clear/stable borders in mountain regions and unclear/unstable borders on the large central European plain. 
We underline that this is another way (popularised in \cite{marshall2016prisoners}) of approaching the problem of culture spreading than what was done in \cite{Dybiec.Mitarai.ea.2012}. 
Although the model is designed in such a way  borders in mountainous regions and along rivers should occur more often, the results show a (somewhat realistic) probabilistic model capturing the naturalistic approach to border formation exists.

% Though, rather than celebrating actual features that we to some extent have put by hand into the model we want to underline that this is a computerisation of another way (popularised in \cite{marshall2016prisoners}) of approaching the problem of culture spreading than in \cite{Dybiec.Mitarai.ea.2012}. 

The model provides a simple framework for thinking about how the geography plays a role in border formation, which to some extent (on a statistical level) captures properties of the interplay between geography and borders. 
Thus, the model could provide some qualitative insight into the influence of geography on the political map of Europe.  

As discussed one could imagine adding additional geographical features, but this would come at the expense of simplicity of the model. 

Another point of discussion in regards to the model is that we have left out the possibility of new ``countries/cultures'' to form. 
With such a possibility one might be able to get a continual dynamics instead of our coarsening dynamics and then study the steady state as was done in \cite{Dybiec.Mitarai.ea.2012}.

One potential way to incorporate the possibility of new countries forming could entail a probability for fracturing of large countries every time step. However, this would introduce more parameters and thus also come at the expense of simplicity of the model.
Additionally, many models of our inspiration in statistical physics (see the review \cite{RevModPhys.81.591}) do not include the possibility of new countries, parties etc. forming. Thus our model could still function as a starting point for such investigations.

\begin{acknowledgments} 
Thanks to Kim Sneppen, Svend Krøjer,  Peter Wildemann, Peter Rasmussen and Kent Bækgaard Lauritsen  for discussions and suggestions. 
FRK acknowledges support from the Villum Foundation for support through the QMATH center of Excellence (Grant No.~10059) and the Villum Young Investigator (Grant No.~25452) programs. 
\end{acknowledgments}

\bibliography{bibliography}

% \begin{appendix*}
\appendix*

\clearpage

\section{Additional plots}\label{sec.app.plots}

%\todo[inline]{Wait with adding errorbars in plots till decision on whether plots go to the final paper is made. }

\subsubsection*{Dependence of country size on radius \texorpdfstring{$R$}{R}}

\begin{figure}[H]
\includegraphics[width=\columnwidth]{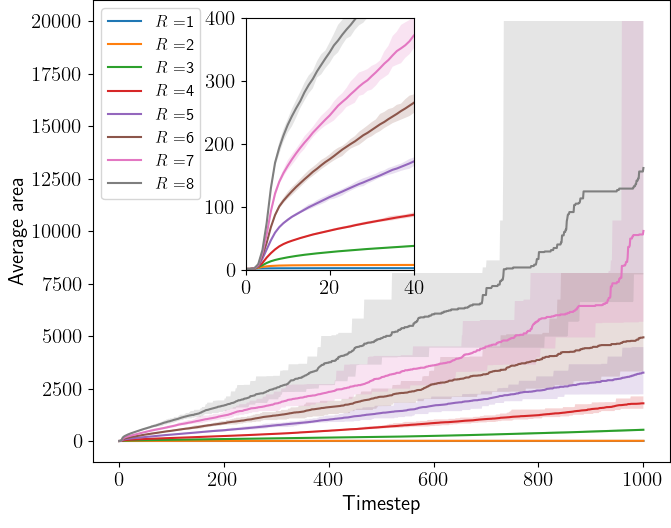}
\caption{Averaged areas of countries over time on the torus (i.e. with periodic boundary conditions) for different values of the parameter $R$ with the fluctuation fixed at $p=0.2$. The results were averaged over 20 iterations 
and the shaded regions are between the $5\%$ and $95\%$ quantiles. We see clearly how the rate of growth of countries increases with $R$ increases.} 
\label{fig:areas_R_torus}
\end{figure}

In \Cref{fig:areas_R_torus} we plot the average size of countries for simulations on the torus for various values of the radius of influence $R$.
As one could expect, the average size of countries increases in as $R$ increases. 
Effectively countries much smaller than $R$ don't have enough local power to defend themselves and get conquered. This leads to a larger average country size for large $R$.

\subsubsection*{Influence of the parameter \texorpdfstring{$A_r$}{Ar}} 
\begin{figure}[H]
\vspace*{1em}
\centering
\includegraphics[width=\columnwidth]{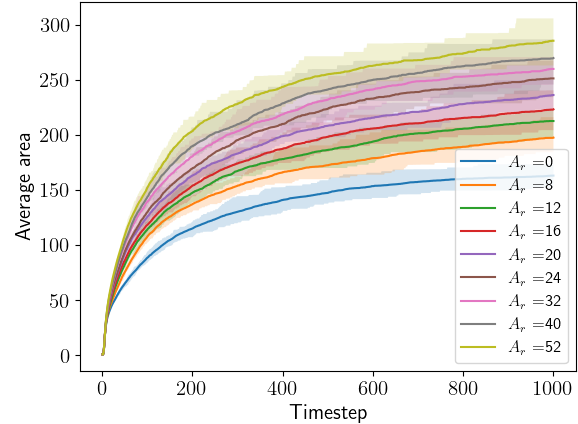}
\caption{Plot of the averaged country size over time for different value of the river area bonus $A_r$. 
Here the parameters were $p=0.2, R=4, D_m=2, P_m=0.6$ instead of the normal choice.
The results were averaged over 20 simulations and the shaded regions are between the $5\%$ and $95\%$ quantiles. The rate of growth of countries increases when the parameter $A_r$ is increased (due to countries forming around rivers). }
\label{parameterAr}
\end{figure}

To study the effect of the parameter $A_r$ we plot in Figure \ref{parameterAr} the average country size against time for different values of the parameter $A_r$.
We see that countries tend to get larger (and larger quicker) when the parameter $A_r$ is large. 
This is the behaviour one would expect, as increased $A_r$ gives countries around rivers more military strength to expand and get larger.

\subsubsection*{Comparison of the two mountain parameters} 
\begin{figure}
    \centering
%\begin{captivy}{
\begin{tabularx}{0.7\linewidth}{*{1}{X}}
\includegraphics[width=0.7\textwidth]{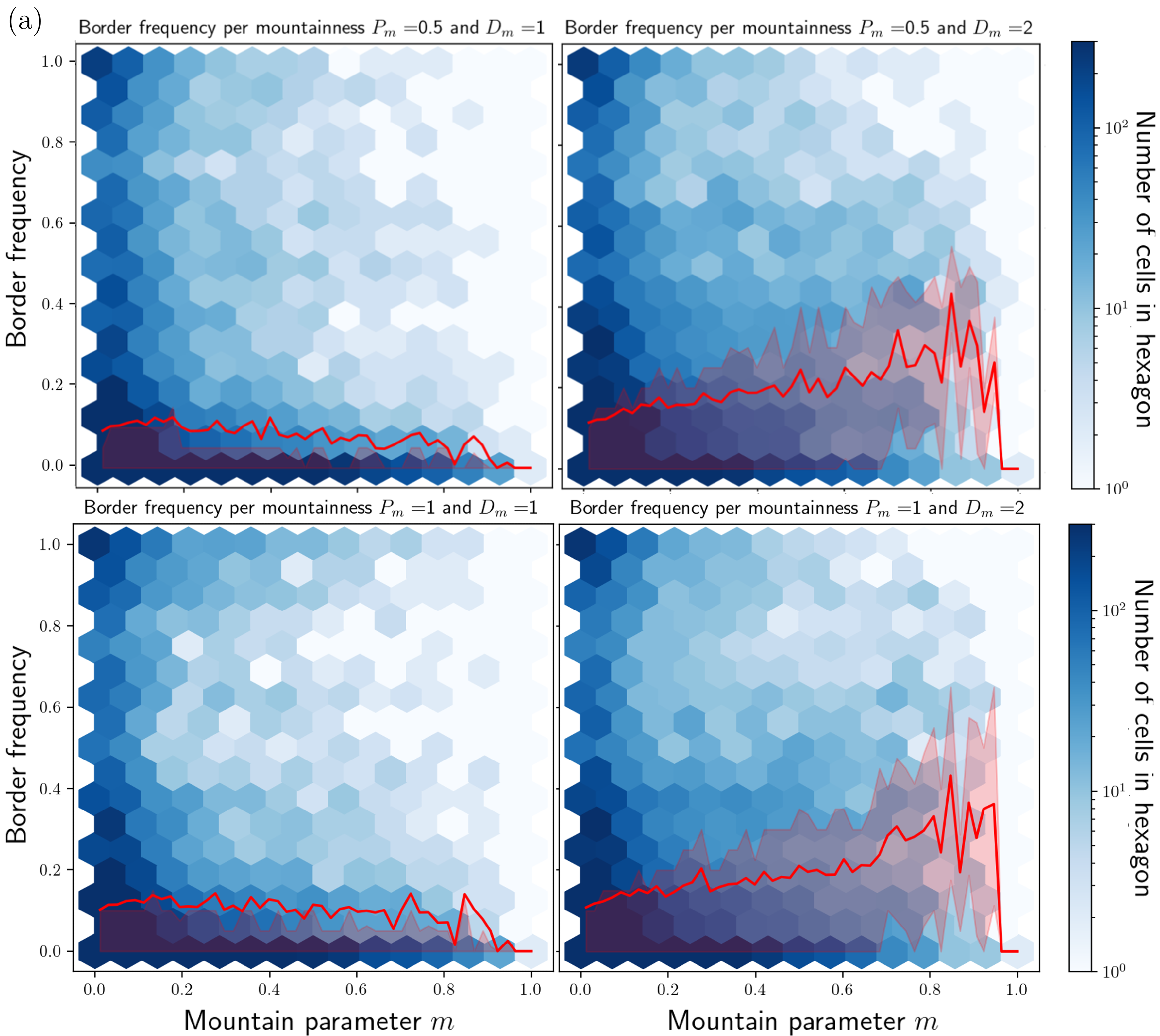}
\\
%\vspace*{1em}
\includegraphics[width=0.7\textwidth]{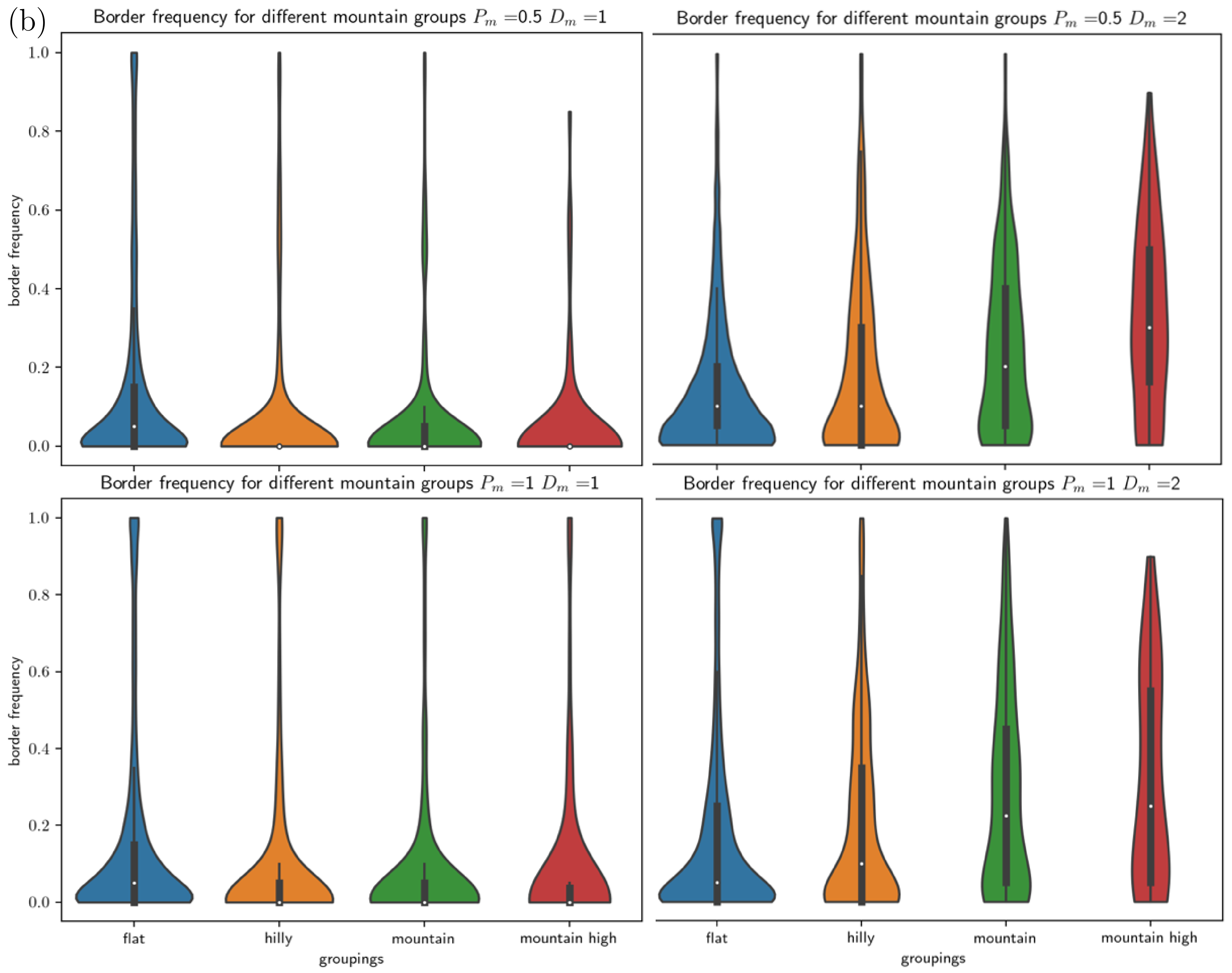}
\end{tabularx}
%}
%\oversubcaption{0.02, 0.98}{}{fig.border.mountain.hexabin}
%\oversubcaption{0.02, 0.45}{}{fig.border.mountain.violin}
%\end{captivy}
\caption{
Distribution of border frequencies for different choices of the parameters $D_m$ and $P_m$. The parameters are chosen to be 
$p = 0.2, R = 4, A_r = 0$ and averaged over 20 iterations. 
%\protect\subref{fig.border.mountain.hexabin} 
(a)
Plot as in \Cref{fig:borders_mountain_simulated}.
The shaded areas denote the 25 and 75 \% quantiles. Note that these error bars indicate the variation of the points with a fixed (binned) mountain parameter and not the variations over iterations.
%\protect\subref{fig.border.mountain.violin} 
(b)
Violin plot.
The category ``flat'' corresponds to the 50 \% land cells with the lowest mountain parameter $m$. The remaining 50 \% are split evenly into the three remaining groups. Note that the two parameters $D_m$ and $P_m$ are clearly different and that $D_m$ is the more important of the two.}
\label{fig.border.mountain.hexabin}
\label{fig.border.mountain.violin}
\end{figure}

We investigate whether the two mountain parameters have different effects. 
To do this we plot the frequency of borders compared to the mountainous parameter $m$.
In \Cref{fig.border.mountain.hexabin}(a) we make a plot as in \Cref{fig:borders_mountain_simulated}. 
In \Cref{fig.border.mountain.violin}(b) we bin all cells into $4$ bins depending on the value $m$. 
Judging from \Cref{fig.border.mountain.hexabin}(a)  it looks like only the mountain defence parameter has a significant effect on the model, but in \Cref{fig.border.mountain.violin}(b) we see that the two mountain parameters have different effects.

\onecolumngrid

\twocolumngrid

\end{document}